\documentclass[manuscript]{aastex62}

\usepackage{txfonts}
\usepackage{latexsym,bm}
\usepackage{url}
\usepackage{color}
\usepackage{colortbl}
\usepackage{multirow}

\definecolor{mygray}{gray}{.9}

\usepackage[titletoc,title]{appendix}

\shorttitle{DIFFUSION WITH ADIABATIC FOCUSING}
\shortauthors{WANG AND QIN}

\begin{document}
              \arraycolsep 0pt

\title{PARALLEL AND PERPENDICULAR DIFFUSION COEFFICIENTS OF ENERGETIC
CHARGED PARTICLES WITH ADIABATIC FOCUSING}

\correspondingauthor{G. Qin}
\email{qingang@hit.edu.cn}

\author[0000-0002-9586-093X]{J. F. Wang}
\affiliation{School of Science, Harbin Institute of Technology, Shenzhen,
518055, China; qingang@hit.edu.cn}

\author[0000-0002-3437-3716]{G. Qin}
\affiliation{School of Science, Harbin Institute of Technology, Shenzhen,
518055, China; qingang@hit.edu.cn}

\begin{abstract}
It is very important
to understand stochastic diffusion of
energetic charged particles
in non-uniform background magnetic field
in plasmas of astrophysics and
fusion devices.
Using different methods
considering along-field adiabatic focusing effect,
various authors derived
parallel diffusion coefficient
$\kappa_\parallel$
and its correction $T$ to
$\kappa_{\parallel 0}$,
where $\kappa_{\parallel 0}$ is
the parallel diffusion coefficient without
adiabatic focusing effect.
In this paper, using the
improved perturbation method developed
by He \& Schlickeiser and iteration process,
we obtain a new
correction $T'$
to $\kappa_{\parallel 0}$.
Furthermore, by employing the
isotropic pitch-angle scattering model
$D_{\mu\mu}=D(1-\mu^2)$,
we find that $T'$
has the different sign as that of $T$.
In this paper the spatial perpendicular
diffusion coefficient $\kappa_\bot$
with the adiabatic focusing effect
is also obtained.

\end{abstract}


\section{INTRODUCTION}

Energetic charged particle propagation
in magnetic turbulent field is
one of the
fundamental problems in astrophysics (e.g.,
cosmic ray physics, astrophysical plasmas, and
space weather research) and
Tokamak fusion devices
\citep[see, e.g.,][]{Jokipii1966,
Schlickeiser2002, MatthaeusEA2003,
ShalchiEA2005,
ShalchiEA2006, Qin2007,
HauffEA2008, Shalchi2009a, Shalchi2010,
QinEA2014}. The magnetic turbulence
can cause the field line wandering,
or the field line random walk (FLRW)
\citep{Jokipii1966, MatthaeusEA1995,
ShalchiAKourakis2007, ShalchiAQin2010,
WangEA2017a}, which
directly affects the diffusion of charged
particles.
In the investigation of
energetic particle transport
through magnetized plasma,
according to observations
one usually assumes
the magnetic field configuration
as the superposition of
a background magnetic field $B_0$
and a turbulent component
$\delta \vec{B}$.
Because the background
magnetic field breaks
the symmetry of the
magnetized plasma,
one have to distinguish
particle diffusion
along and across
the large-scale magnetic field.
However, some previous articles
only consider the parallel
diffusion, since it is
much greater than the perpendicular
one in many scenarios
\citep[see, e.g.,][]{Earl1974, Earl1976,
BeeckEA1986, Shalchi2011,
Litvinenko2012a,
Litvinenko2012b, ShalchiEA2013, HeEA2014}.
On the other hand, in some conditions, e.g.,
with high turbulence levels,
perpendicular diffusion of
energetic particles becomes too important
to be ignored
\citep{DwyerEA1997, ZhangEA2003}.

Various analytical theories of parallel and
perpendicular diffusion for energetic
charged particles have been
developed in the past.
\citet{Jokipii1966} established the
quasi-linear theory of
diffusion which corresponds to the first order
perturbation theory.
However, the quasi-linear theory
is problematic and
nonlinear description for particle
propagation is essential, especially
for particle perpendicular diffusion
\citep[e.g.,][]{QinEA02}.
NonLinear Guiding Center (NLGC)
theory developed by
\citet{MatthaeusEA2003} is the
important breakthrough,
which agrees well with
the computer simulations
of two-component
(slab+two-dimensional)
turbulence model
\citep[see,][]{MatthaeusEA90}.
Based on NLGC, \citet{Shalchi2010}
developed the Unified NonLinear
Transport (UNLT)
theory for perpendicular
diffusion coefficient, using the
Fokker-Planck equation
of energetic charged particles to
treat the fourth-order
correlation. It is suggested that
UNLT can be applied
for arbitrary turbulence
geometry to explain the
subdiffusive transport
for slab turbulence and the
recovery of diffusion
for three-dimensional and
quasi-three-dimensional
turbulence models.

By radio continuum surveys of
interstellar space and direct
in-situ measurements
in solar system, it is well established
that for many scenarios the background
magnetic fields are spatially varying.
However, the above research
about parallel and perpendicular
diffusion only explored
the uniform mean magnetic field.
One can show that the spatially varying
background magnetic fields lead
to the adiabatic focusing effect of
charged energetic particle
transport and introduces
correction to the particle
diffusion coefficients
\citep[see, e.g.,][]{Roelof1969,
Earl1976, Kunstmann1979,BeeckEA1986,
BieberEA1990,
Kota2000, SchlickeiserEA2008,
Shalchi2009b, Shalchi2011, Litvinenko2012a,
Litvinenko2012b, ShalchiEA2013, WangEA2016,
WangEA2017b}. To explore the influence of
adiabatic focusing on particle transport,
perturbation method is frequently used
\citep[see, e.g.,][]{BeeckEA1986,
BieberEA1990, SchlickeiserEA2008,
SchlickeiserEA2010,
LitvinenkoASchlickeiser2013, HeEA2014}.
To use the perturbation method, since
anisotropic distribution function
is an implicit function,
by using iteration method
one can find that the anisotropic
distribution function
becomes infinite series of the
spatial and temporal derivatives
of the isotropic distribution
function. Therefore, the governing
equation of isotropic distribution
function derived from the Fokker-Planck
equation contains infinite terms
because of the infinite series of
anisotropic distribution function.
By using
truncating method to neglect the
higher-order derivative terms,
the approximate correction formulas
of parallel or perpendicular
diffusion coefficients were obtained
\citep[see, e.g.,][]
{SchlickeiserEA2008,
SchlickeiserEA2010, HeEA2014}.
However, the higher-order
derivative terms
probably also make the correction to the
parallel and perpendicular diffusion
as the lower-order derivative ones.
Maybe the magnitude of the correction
from the higher-order derivative terms
might not be
the higher order small quantity than
that of the correction
from the lower-order derivative terms.
Therefore, the correction obtained
by the previous authors
is likely to contain significant error.
In this paper by considering
the higher-order derivative terms
we derive the parallel
and perpendicular diffusion
coefficients and obtain the
correction formulas
coming from all order derivative terms
by using
the improved perturbation method
\citep{HeEA2014}
and the iteration operation.
And for the weak adiabatic
focusing limit
we evaluate the correction to the
parallel diffusion coefficient
and compare it
with the correction obtained
in the previous papers.

The paper is organized as follows.
In Section \ref{EQUATION OF ISOTROPIC
DISTRIBUTION FUNCTION}, by considering
adiabatic focusing effect, we derive
the governing equation of isotropic
distribution function with the infinite
series. In Section \ref{ANALYTICAL
COEFFICIENTS FOR Lambda=0}, by employing
the truncation, we deduce
the approximate formulas of the
perpendicular and parallel diffusion
coefficients, and that of the streaming
term.
In Section \ref{ANALYTICAL COEFFICIENTS
FOR Lambda ne 0}, the parallel and
perpendicular diffusion coefficients
$\kappa_\parallel$ and $\kappa_\bot$,
and the parallel streaming
coefficient $\kappa_1$ are derived,
with the influence of the infinite
series caused by the iteration of
anisotropic distribution function.
We conclude and summarize
our results in Section
\ref{SUMMARY AND CONCLUSION}.

\section{EQUATION OF ISOTROPIC
DISTRIBUTION FUNCTION}
\label{EQUATION OF ISOTROPIC
DISTRIBUTION FUNCTION}

The starting point of this paper is
the modified Fokker-Planck equation
for the gyro-tropic energetic charged
particle distribution function, which
incorporates the pitch-angle and
perpendicular diffusion, and the
along-field adiabatic focusing
\begin{equation}
\frac{\partial{f}}{\partial{t}}
+ v\mu
\frac{\partial{f}}
{\partial{z}}=
\frac{\partial{}}{\partial{\mu}}
\left[D_{\mu \mu}\frac{\partial{f}}
{\partial{\mu}}-\frac{v}{2L}
(1-\mu^2)f \right]
+D_\bot  \Delta_\bot f.
\label{Fokker-Planck equation}
\end{equation}
Here $t$ is time, $z$
is the distance
along the background magnetic field,
$\mu=v_z /v$ is the pitch-angle
cosine with particle speed $v$ and
its z-component $v_z$, $D_{\mu \mu}$
is the pitch-angle diffusion
coefficient, $D_\bot$ is the
Fokker-Planck perpendicular
diffusion coefficient,
$L(z)=-B_0 (z)/ [dB_0 (z) / dz]$
is the adiabatic focusing
characteristic length of the
large-scale magnetic field $B_0(z)$,
and $\Delta_\bot=\partial^2{} /
\partial{x^2}+\partial^2{}
/ \partial{y^2}$
is the differential operator across
the large-scale magnetic field.
The source term is not included
in the above equation.
Because only the influence
of the along-field
adiabatic focusing
on the parallel and
perpendicular diffusion
is explored in this paper,
the terms related to
momentum diffusion
and so on are ignored
in the Fokker-Planck Equation
(\ref{Fokker-Planck equation}).
The more complete form of
the Fokker-Planck equation
can be found in
\citet{Schlickeiser2002}.

It should be mentioned that
the linear phase space density
$f(x, y, z, p, \mu, t)
=f_0(x, y, z, p, \mu, t)/B_0(z)$,
so that Equation
(\ref{Fokker-Planck
equation}) is equivalent to the
standard Fokker-Planck equation
\begin{equation}
\frac{\partial{f_0}}{\partial{t}}
+ v\mu
\frac{\partial{f_0}}
{\partial{z}}
=\frac{\partial{}}
{\partial{\mu}}
\left[D_{\mu \mu}
\frac{\partial{f_0}}
{\partial{\mu}}\right]
-\frac{v}{2L}(1-\mu^2)
\frac{\partial{f_0 }}
{\partial{\mu}}
+D_\bot  \Delta_\bot f_0.
\label{standard Fokker-Planck
equation}
\end{equation}

If the pitch-angle scattering
is strong,
the gyro-tropic phase space
distribution
function quickly becomes
quasi-equilibrium
distribution. Therefore,
we can split the gyro-tropic
cosmic-ray phase space
density $f(\vec{x}, \mu, t)$
into the dominant isotropic part
$F(\vec{x}, t)$
and the subordinate anisotropic part
$g(\vec{x}, \mu, t)$
as following the previous articles
\citep[see, e.g.,][]
{SchlickeiserEA2007,
SchlickeiserEA2008, HeEA2014}
\begin{equation}
f(\vec{x}, \mu, t)=F(\vec{x}, t)
+g(\vec{x}, \mu, t)
\end{equation}
with
\begin{equation}
F(\vec{x}, t)=\frac{1}{2}
\int_{-1}^1 d\mu
f(\vec{x}, \mu, t)
\end{equation}
and
\begin{equation}
\int_{-1}^1 d\mu
g(\vec{x}, \mu, t)=0 .
\end{equation}

\subsection{The differential
equation
of anisotropic distribution
function $g(\mu)$}
\label{The differential equation
of anisotropic distribution
function g(mu)}

In this subsection, we would adopt
the method in \citet{HeEA2014} to
derive
the differential equation of the
anisotropic distribution function
$g(\mu)$.

By integrating Equation
(\ref{Fokker-Planck equation})
over $\mu$ from -1 to 1
we can obtain
\begin{equation}
\frac{\partial{F}}{\partial{t}}
+ \frac{v}{2}
\frac{\partial{}}
{\partial{z}}\int_{-1}^{1}
\mu g d\mu=
\frac{1}{2}\Delta_\bot F
\int_{-1}^{1}d\mu D_\bot
+ \frac{1}{2}\Delta_\bot
\int_{-1}^{1}d\mu D_\bot g.
\label{Equation of F with g}
\end{equation}
Furthermore, by integrating
Equation
(\ref{Fokker-Planck equation})
over $\mu$ from -1 to $\mu$,
the following equation
can be found
\begin{eqnarray}
\frac{\partial{F}}
{\partial{t}}(\mu+1)
&&+ \frac{\partial{}}
{\partial{t}}
\int_{-1}^{\mu}d\mu g
+\frac{v(\mu^2-1)}{2}
\frac{\partial{F}}
{\partial{z}}
+v\frac{\partial{}}
{\partial{z}}
\int_{-1}^{\mu}
d\mu \mu g \nonumber\\
&&=D_{\mu\mu}
\frac{\partial{g}}
{\partial{\mu}}
-\frac{v(1-\mu^2)}{2L}F
-\frac{v(1-\mu^2)}{2L}g
+\Delta_\bot\int_{-1}^
{\mu}d\mu D_\bot F+
\Delta_\bot\int_{-1}^{\mu}
d\mu D_\bot g.
\label{integrate
from -1 to mu}
\end{eqnarray}
To obtain the Equations
(\ref{Equation of F with g})
and (\ref{integrate from -1
to mu}) the assumption
$D_{\mu\mu}(\mu=\pm 1)=0$
is used.
Subtracting Equation
(\ref{Equation of F with g})
from
(\ref{integrate from -1 to mu})
we can get
\begin{equation}
\frac{\partial{g}}
{\partial{\mu}}-
\frac{v(1-\mu^2)g}
{2L D_{\mu \mu}}
+\frac{v(1-
\mu^2)}{2D_{\mu \mu}}
\left(\frac{\partial{F}}
{\partial{z}}
-\frac{F}
{L}\right)=\Phi(\mu)
\label{Equation of g with phi}
\end{equation}
with
\begin{eqnarray}
\Phi(\mu)=&&\frac{1}{D_{\mu\mu}}
\Bigg[\left(\frac{\partial{F}}
{\partial{t}}\mu
+\frac{\partial{}}{\partial{t}}
\int_{-1}^{\mu}gd\nu\right)
-\Delta_\bot \left(\int_{-1}^{\mu}
d\nu D_\bot g-\frac{1}{2}
\int_{-1}^{1}d\mu D_\bot g\right)
\nonumber   \\
&+&\frac{v}{2}\frac{\partial{}}
{\partial{z}}
\left(2\int_{-1}^{\mu}d\nu \nu g-
\int_{-1}^{1}d\mu \mu g\right)
-\Delta_\bot F
\left(\int_{-1}^{\mu}d\nu D_\bot-
\frac{1}{2}\int_{-1}^{1}d\mu
D_\bot \right)\Bigg].
\label{Phi}
\end{eqnarray}
By defining the following
quantity
\begin{eqnarray}
M(\mu)&=&\frac{v}{2L}
\int_{-1}^{\mu} d\nu
\frac{1-\nu^2}{D_{\mu \mu}(\nu)},
\label{M(mu)}
\end{eqnarray}
we can obtain
\begin{equation}
\frac{\partial{}}{\partial{\mu}}
\Bigg\{\Bigg[g(\mu)
-L\left(\frac{\partial{F}}
{\partial{z}}
-\frac{F}{L} \right)
\Bigg]e^{-M(\mu)}\Bigg\}
=e^{-M(\mu)}\Phi(\mu).
\label{HS-like}
\end{equation}
It is noted that in \citet{HeEA2014}
the right hand side of
this equation (Equation (20) in
their paper) is set to be $0$.

\subsection{The integration
results of $g(\mu)$}
\label{The integration results
of g(mu)}

Through integrating
Equation (\ref{HS-like})
over $\mu$,
we can get the anisotropic
distribution function
as the following
\begin{equation}
g(\mu)=L\left(\frac{\partial{F}}
{\partial{z}}
-\frac{F}{L}\right)\left[1-
\frac{2e^{M(\mu)}}{\int_{-1}^{1}
d\mu e^{M(\mu) }}\right]
+e^{M(\mu)}\left[R(\mu)
-\frac{\int_{-1}^{1}d\mu
e^{M(\mu)}R(\mu)}
{\int_{-1}^{1}d\mu
e^{M(\mu) }}\right]
\label{g}
\end{equation}
with
\begin{eqnarray}
R(\mu)&=&\int_{-1}^{\mu} d\nu
e^{-M(\nu)}\Phi(\nu).
\label{R(mu)}
\end{eqnarray}
Equation (\ref{g}) contains
the effect from the term on the
right-hand side of
Equation (\ref{HS-like}),
whereas the Equation (20)
in \citet{HeEA2014}
is an approximate expression.

Inserting $R(\mu)$
(Equation (\ref{R(mu)}))
into $g(\mu)$ (Equation (\ref{g}))
and considering
Equation (\ref{Phi}),
we find that the anisotropic
distribution
function $g(\mu)$ becomes
\begin{eqnarray}
g(\mu)&=&L\left(\frac{\partial{F}}
{\partial{z}}
-\frac{F}{L}\right)\left[1-
\frac{2e^{M(\mu)}}{\int_{-1}^{1}
d\mu e^{M(\mu) }}\right]
+e^{M(\mu)}\Bigg\{\int_{-1}^{\mu}
d\nu
e^{-M(\nu)}\frac{1}{D_{\mu\mu}}
\Bigg[\left(\frac{\partial{F}}
{\partial{t}}\mu
+\frac{\partial{}}{\partial{t}}
\int_{-1}^{\mu}gd\nu\right)
\nonumber \\
&&-\Delta_\bot
\left(\int_{-1}^{\mu}
d\nu D_\bot g-\frac{1}{2}
\int_{-1}^{1}d\mu D_\bot g\right)
+\frac{v}{2}\frac{\partial{}}
{\partial{z}}
\left(2\int_{-1}^{\mu}d\nu \nu g-
\int_{-1}^{1}d\mu \mu g\right)
-\Delta_\bot F
\left(\int_{-1}^{\mu}d\nu D_\bot-
\frac{1}{2}\int_{-1}^{1}d\mu
D_\bot \right)\Bigg]\nonumber\\
&&-\frac{1}
{\int_{-1}^{1}d\mu
e^{M(\mu) }}\int_{-1}^{1}d\mu
e^{M(\mu)}\int_{-1}^{\mu} d\nu
e^{-M(\nu)}\frac{1}{D_{\mu\mu}}
\Bigg[\left(\frac{\partial{F}}
{\partial{t}}\mu
+\frac{\partial{}}{\partial{t}}
\int_{-1}^{\mu}gd\nu\right)
-\Delta_\bot \left(\int_{-1}^{\mu}
d\nu D_\bot g-\frac{1}{2}
\int_{-1}^{1}d\mu D_\bot g\right)
\nonumber\\
&&+\frac{v}{2}\frac{\partial{}}
{\partial{z}}
\left(2\int_{-1}^{\mu}d\nu \nu g-
\int_{-1}^{1}d\mu \mu g\right)
-\Delta_\bot F
\left(\int_{-1}^{\mu}d\nu D_\bot-
\frac{1}{2}\int_{-1}^{1}d\mu
D_\bot \right)\Bigg]\Bigg\}.
\label{g2}
\end{eqnarray}
From the latter equation
we can find that
the anisotropic distribution
function $g(\mu)$
is an implicit function.
Therefore,
by iterating Equation (\ref{g2}),
i.e. applying it
repeatedly,
we obtain the formula of
$g(\mu)$ as
\begin{equation}
g(\mu)=\sum_{m, n, p}
\epsilon_{m,n,p}
\frac{\partial^{m+n}{}}
{\partial {t^m}
\partial{z}^n}\Delta_\bot^p  F
\label{expanded g}
\end{equation}
with the coefficients
$\epsilon_{m,n,p}$.
Here $m,n,p=0,1,2,3,\cdots$.
Similarily,
by inserting Equation (\ref{g})
into Equation (\ref{Phi})
and inputting Equation (\ref{Phi})
into Equation (\ref{R(mu)}),
we can find that
$R(\mu)$ is also an implicit
 function. Therefore,
 with iteration operation
$R(\mu)$ can also
be written as
\begin{equation}
R(\mu)=\sum_{m, n, p}
\chi_{m,n,p}
\frac{\partial^{m+n}{}}
{\partial {t^m}
\partial{z}^n}\Delta_\bot^p  F
\label{series of R}
\end{equation}
with the coefficients
$\chi_{m,n,p}$ and
$m,n,p=0,1,2,3,\cdots$.
Apparently, the coefficients
$\chi_{m,n,p}$
in the latter equation
is related to the coefficients
$\epsilon_{m,n,p}$ in Equation
(\ref{expanded g})
since $g(\mu)$
and $R(\mu)$ are
related according
to Equations (\ref{Phi}), (\ref{g}),
(\ref{R(mu)}), and (\ref{g2}).

\subsection{The governing
equation of the
isotropic distribution function
$F(\vec{x}, t)$}
\label{The governing equation
of the
isotropic distribution function
F}

From Equation
(\ref{Equation of F with g})
we can find that in order to
obtain the governing equation
of isotropic distribution
function
$F(\vec{x}, t)$ we have to get
the expression of
$\int_{-1}^{1}d\mu \mu g$ and
$\int_{-1}^{1}d\mu  D_\bot g$.
By using Equation (\ref{g})
we can obtain
\begin{equation}
\int_{-1}^{1}d\mu \mu g
=-2\frac{\int_{-1}^{1}d\mu\mu
e^{M(\mu)}}
{\int_{-1}^{1}d\mu e^{M(\mu)}}
\left(\frac{\partial{F}}
{\partial{z}}
-\frac{F}{L}\right)L+\int_{-1}^{1}
d\mu \mu e^{M(\mu)}
\Bigg[R(\mu)
-\frac{\int_{-1}^{1}d\mu
e^{M(\mu)}R(\mu)}
{\int_{-1}^{1}d\mu e^{M(\mu)}}
\Bigg]
\label{Integrating mu g over
mu from -1 to 1}
\end{equation}
and
\begin{equation}
\int_{-1}^{1}d\mu  D_\bot g
=L\left(\frac{\partial{F}}
{\partial{z}}
-\frac{F}{L}\right)\int_{-1}^{1}
d\mu D_\bot
\left[1-\frac{2 e^{M(\mu)}}
{\int_{-1}^{1}d\mu
e^{M(\mu)}} \right]
+\int_{-1}^{1}d\mu D_\bot
e^{M(\mu)}
\left[R(\mu)
-\frac{\int_{-1}^{1}d\mu
e^{M(\mu)}R(\mu)}{\int_{-1}^{1}
d\mu e^{M(\mu)}}\right].
\label{Integrating
D perpendicular
g over mu from -1 to 1}
\end{equation}

Substituting Equations
(\ref{Integrating mu g over
mu from -1 to 1})
and
(\ref{Integrating D
perpendicular g
over mu from -1 to 1})
into Equation
(\ref{Equation of F with g})
yields
\begin{eqnarray}
\frac{\partial{F}}{\partial{t}}
&-&\frac{\partial{}}
{\partial{z}}
\left[vL\frac{\int_{-1}^{1}
d\mu \mu e^{M(\mu)}}
{\int_{-1}^{1}d\mu
e^{M(\mu)}}\left(\frac{\partial{F}}
{\partial{z}}-
\frac{F}{L}\right)\right]
-\Delta_\bot F \frac{\int_{-1}^{1}
d\mu D_\bot e^{M(\mu)}}
{\int_{-1}^{1}d\mu e^{M(\mu)}}
\nonumber\\
&-&\frac{L}{2}\Delta_\bot
\frac{\partial{F}}{\partial{z}}
\int_{-1}^{1}d\mu D_\bot
\left[1-\frac{2 e^{M(\mu)}}
{\int_{-1}^{1}
d\mu e^{M(\mu)}} \right]
=\Lambda(x,y,z,t)
\label{Accurate diffusion equation}
\end{eqnarray}
with
\begin{equation}
\Lambda(x,y,z,t)=
-\frac{v}{2}\int_{-1}^{1}
d\mu \mu e^{M(\mu)}
\left[\frac{\partial{R}}
{\partial{z}}
-\frac{\int_{-1}^{1}d\mu
\frac{\partial{R}}{\partial{z}}
e^{M(\mu)}}
{\int_{-1}^{1}d\mu e^{M(\mu)}}
\right]+
\frac{1}{2}\int_{-1}^{1}
d\mu D_\bot
e^{M(\mu)}\left[\Delta_\bot R-
\frac{\int_{-1}^{1}
d\mu e^{M(\mu)}
\Delta_\bot R}
{\int_{-1}^{1}
d\mu e^{M(\mu)}} \right].
\label{Lambda}
\end{equation}
By inserting Equation
(\ref{series of R})
into the latter equation
we can obtain the
following formula
\begin{equation}
\Lambda(x,y,z,t)
=\sum_{m,n,p}\eta_{m,n,p}
\frac{\partial^{m+n}}
{\partial {t^m}
\partial{z}^n}
\Delta_\bot^p  F
\label{Lambda-eta}
\end{equation}
with the coefficients
$\eta_{m,n,p}$ and
$m,n,p=0,1,2,3,\cdots$.
Note that some of the
coefficients
$\eta_{m,n,p}$ are equal to 0.

Replacing $\Lambda(x,y,z,t)$ in
Equation (\ref{Accurate diffusion
equation}) with the
latter equation
gives
\begin{eqnarray}
\frac{\partial{F}}{\partial{t}}
-\frac{\partial{}}
{\partial{z}}&&
\left[vL\frac{\int_{-1}^{1}
d\mu \mu e^{M(\mu)}}
{\int_{-1}^{1}d\mu
e^{M(\mu)}}
\left(\frac{\partial{F}}
{\partial{z}}-
\frac{F}{L}\right)\right]
-\Delta_\bot F
\frac{\int_{-1}^{1}
d\mu D_\bot e^{M(\mu)}}
{\int_{-1}^{1}d\mu e^{M(\mu)}}
\nonumber\\
&-&\frac{L}{2}\Delta_\bot
\frac{\partial{F}}{\partial{z}}
\int_{-1}^{1}d\mu D_\bot
\left[1-\frac{2 e^{M(\mu)}}
{\int_{-1}^{1}
d\mu e^{M(\mu)}} \right]
=\sum_{m,n,p}\eta_{m,n,p}
\frac{\partial^{m+n}}
{\partial {t^m}
\partial{z}^n}
\Delta_\bot^p  F.
\label{Accurate diffusion
equation-2}
\end{eqnarray}
The latter equation is
the most general
resulting transport equation
in this paper.

After combining similar terms
for the latter equation,
we can obtain
\begin{eqnarray}
(\eta_{1,0,0}+1)
\frac{\partial{F}}{\partial{t}}
&+&\left(\eta_{0,1,0}
+v\frac{\int_{-1}^{1}d\mu
\mu e^{M(\mu)}}
{\int_{-1}^{1}
d\mu e^{M(\mu)}}\right)
\frac{\partial{F}}{\partial{z}}
=\left(\eta_{0,2,0}
+vL\frac{\int_{-1}^{1}d\mu
\mu e^{M(\mu)}}
{\int_{-1}^{1}d\mu
e^{M(\mu)}}\right)
\frac{\partial^2{F}}
{\partial{z}^2}
+\left(\eta_{0,0,1}
+\frac{\int_{-1}^{1}d\mu D_\bot
e^{M(\mu)}}{\int_{-1}^{1}
d\mu e^{M(\mu)}}\right)
\Delta_\bot F \nonumber\\
&+&\left(\eta_{0,1,1}
+\frac{L}{2}\int_{-1}^{1}
d\mu D_\bot
\left[1-\frac{2e^{M(\mu)}}
{\int_{-1}^{1}d\mu
e^{M(\mu)}}\right]\right)
\Delta_\bot \frac{\partial{F}}
{\partial{z}}
+\sum_{(m,n,p)\notin A}
\eta_{m,n,p}
\frac{\partial^{m+n}}
{\partial {t^m}
\partial{z}^n}
\Delta_\bot^p  F.
\label{Accurate diffusion
equation-CST}
\end{eqnarray}
Here, $A=\{(1,0,0), (0,1,0),
(0,2,0), (0,0,1), (0,1,1)\}$.

Furthermore, Equation
(\ref{Accurate diffusion
equation-CST})
can be simply rewritten as
\begin{eqnarray}
\kappa_{1,0,0}\frac{\partial{F}}
{\partial{t}}
&+&\kappa_{0,1,0}
\frac{\partial{F}}{\partial{z}}
=\sum_{(m,n,p)\notin B}\kappa_{m,n,p}
\frac{\partial^{m+n}}
{\partial {t^m}
\partial{z}^n}
\Delta_\bot^p  F
\label{most general equation
of isotropic distribution function}
\end{eqnarray}
with
\begin{equation}
\kappa_{m,n,p}=\left\{
\begin{array}
{l@{\quad \quad}l}
\eta_{1,0,0}+1,
& (m,n,p)=(1,0,0) \\
\eta_{0,1,0}
+v\frac{\int_{-1}^{1}d\mu
\mu e^{M(\mu)}}
{\int_{-1}^{1}d\mu e^{M(\mu)}},
& (m,n,p)=(0,1,0) \\
\eta_{0,0,1}+\frac{\int_{-1}^{1}
d\mu D_\bot
e^{M(\mu)}}{\int_{-1}^{1}
d\mu e^{M(\mu)}},&
(m,n,p)=(0,0,1)\\
\eta_{0,2,0}
+vL\frac{\int_{-1}^{1}d\mu
\mu e^{M(\mu)}}{\int_{-1}^{1}
d\mu e^{M(\mu)}},
& (m,n,p)=(0,2,0) \\
\eta_{0,1,1}+\frac{L}{2}
\int_{-1}^{1}
d\mu D_\bot
\left[1-\frac{2e^{M(\mu)}}
{\int_{-1}^{1}
d\mu e^{M(\mu)}}\right],
& (m,n,p)=(0,1,1)\\
\eta_{m,n,p} &\text{otherwise}.
\end{array}\right.
\end{equation}
Here, $B=\{(1,0,0), (0,1,0)\}$,
 $\kappa_{1,0,0}$
is the coefficient
of the first order
time derivative term,
$\kappa_{0,1,0}$
is the coefficient
of parallel streaming term,
$\kappa_{0,2,0}$
is the parallel
diffusion coefficient,
$\kappa_{0,0,1}$
is the perpendicular
diffusion coefficient,
and $\kappa_{0,1,1}$
is the coefficient
of the
term with
$\Delta_\bot\partial{F}/\partial{z}$.
Equation (\ref{Accurate diffusion
equation-CST}) and
(\ref{most general equation
of isotropic
distribution function})
are equivalent to Equation
(\ref{Accurate diffusion
equation-2}). They are all the forms
of the governing equation of the
isotropic distribution function
in the most general case.

In this paper, we only explore
the properties of
the first order time derivative term
coefficient $\kappa_{1,0,0}$,
the parallel streaming coefficient
$\kappa_{0,1,0}$,
the parallel diffusion coefficient
$\kappa_{0,2,0}$, and
the perpendicular
diffusion coefficient
$\kappa_{0,0,1}$, which
are listed as
\begin{eqnarray}
\kappa_{1,0,0}&=&\eta_{1,0,0}
+\kappa_{1,0,0}^a,
\label{k100}\\
\kappa_{0,1,0}&=&\eta_{0,1,0}
+\kappa_{0,1,0}^a,
\label{k010}\\
\kappa_{0,0,1}&=&\eta_{0,0,1}
+\kappa_{0,0,1}^a
\label{k001}\\
\kappa_{0,2,0}&=&\eta_{0,2,0}
+\kappa_{0,2,0}^a,
\label{k020}
\end{eqnarray}
with
\begin{eqnarray}
\kappa_{1,0,0}^a&=&1,\\
\kappa_{0,1,0}^a &=&
v\frac{\int_{-1}^{1}d\mu
\mu e^{M(\mu)}}
{\int_{-1}^{1}
d\mu e^{M(\mu)}},  \\
\kappa_{0,0,1}^a&=&
\frac{\int_{-1}^{1}d\mu D_\bot
e^{M(\mu)}}{\int_{-1}^{1}
d\mu e^{M(\mu)}},\\
\kappa_{0,2,0}^a &=&
vL\frac{\int_{-1}^{1}d\mu
\mu e^{M(\mu)}}{\int_{-1}^{1}
d\mu e^{M(\mu)}}.
\end{eqnarray}
Note that any of the coefficients
$\eta_{1,0,0}$, $\eta_{0,1,0}$,
$\eta_{0,0,1}$, and $\eta_{0,2,0}$
might be zero.

\section{ANALYTICAL COEFFICIENTS
WITH $\Lambda(x,y,z,t)= 0$}
\label{ANALYTICAL COEFFICIENTS
FOR Lambda=0}

For the condition
$\Lambda(\vec{x}, \mu, t)=0$,
from Equation (\ref{Lambda})
we can find that $R(\mu)=0$.
So, from formula (\ref{g})
the approximate
anisotropic distribution function
can be obtained
\begin{equation}
g^a(\mu)=L\left(\frac{\partial{F}}
{\partial{z}}-\frac{F}{L}\right)
\left[1-\frac{2e^{M(\mu)}}
{\int_{-1}^{1}
d\mu e^{M(\mu) }}\right],
\label{approximate formula of g}
\end{equation}
which is identical with
the Equation (20) in
\citet{HeEA2014}.
In addition, by using
the condition
$\Lambda(\mu)=0$,
i.e., setting $\eta_{m,n,p}=0$
for any $m,n,p$,
we can simplify
Equation
(\ref{Accurate diffusion
equation-CST})
as
\begin{equation}
\frac{\partial{F}}{\partial{t}}
+\kappa_1^a\frac{\partial{F}}
{\partial{z}}
=\frac{\partial{}}{\partial{z}}
\left(\kappa_\parallel^a
\frac{\partial{F}}{\partial{z}}
\right)+\kappa_\bot^a
\Delta_\bot F+\kappa_3^a
\Delta_\bot
\frac{\partial{F}}{\partial{z}}
\label{Diffusion equation with R=0}
\end{equation}
with
\begin{eqnarray}
\kappa_t^a&=&\kappa_{1,0,0}^a=1,
\label{Approximate time derivative}\\
\kappa_1^a &=&\kappa_{0,1,0}^a=
\frac{\kappa_\parallel}{L}
=v\frac{\int_{-1}^{1}d\mu
\mu e^{M(\mu)}}
{\int_{-1}^{1}d\mu e^{M(\mu)}}     \\
\kappa_\bot^a &=& \kappa_{0,0,1}^a=
\frac{\int_{-1}^{1}d\mu D_\bot
e^{M(\mu)}}{\int_{-1}^{1}
d\mu e^{M(\mu)}}
\label{Perpendicular diffusion
coefficient for phi=0}, \\
\kappa_\parallel^a&=&\kappa_{0,2,0}^a
=vL\frac{\int_{-1}^{1}d\mu
\mu e^{M(\mu)}}{\int_{-1}^{1}
d\mu e^{M(\mu)}}
\label{Approximate spatial parallel
diffusion coefficient}, \\
\kappa_3^a &=&\frac{L}{2}
\int_{-1}^{1}
d\mu D_\bot \left[1
-\frac{2e^{M(\mu)}}
{\int_{-1}^{1}d\mu
e^{M(\mu)}}\right].
\label{Approximate k3}
\end{eqnarray}
Obviously, Equations
(\ref{Approximate time derivative})-(\ref{Approximate k3})
are the special forms of
Equations (\ref{k100})-(\ref{k001})
with
$\Lambda(x,y,z,t)=0$.
In particular, Equation
(\ref{Perpendicular diffusion
coefficient for phi=0}) shows
the approximate perpendicular
diffusion coefficient
with the adiabatic focusing effect.
Furthermore, the approximate parallel
diffusion coefficient
$\kappa_\parallel^a$
(Equation
(\ref{Approximate spatial parallel
diffusion coefficient})) is identical
with the result obtained by some
previous authors \citep[see,][]
{BeeckEA1986, Litvinenko2012a,
HeEA2014}, and it can be written as
\begin{equation}
\kappa_\parallel^a
=\kappa_{\parallel 0}+T,
\label{T}
\end{equation}
where $T$ is the correction to
$\kappa_{\parallel0}$.

\section{ANALYTICAL COEFFICIENTS
WITH $\Lambda(x,y,z,t)\ne 0$}
\label{ANALYTICAL COEFFICIENTS
FOR Lambda ne 0}

In general, if
$\Lambda(x,y,z,t)\ne 0$,
we have to obtain the
coefficient formulas of
the isotropic distribution
function equation (see Equation
(\ref{Accurate diffusion
equation}))
with the influence of
$\Lambda(\mu)$.
Therefore, we need to get the
terms $\partial{R} / \partial{z}$ and
$\Delta_\bot R$ because of
the Equation (\ref{Lambda}),
in consequence, we have to
get the formula for $R(\mu)$.
By combining Equations (\ref{Phi})
and (\ref{R(mu)})
we can get the following formula
\begin{eqnarray}
R(\mu)=&&\int_{-1}^{\mu}d\nu
\frac{e^{-M(\nu)}}
{D_{\mu \mu}}
\Bigg[\left(\frac{\partial{F}}
{\partial{t}}\nu+\frac{\partial{}}
{\partial{t}}
\int_{-1}^{\nu}d\rho g\right)
+\frac{v}
{2}\Bigg(2\int_{-1}^{\nu}d\rho
\rho \frac{\partial{g}}
{\partial{z}}
-\int_{-1}^{1}d\mu \mu
\frac{\partial{g}}{\partial{z}}
\Bigg)\nonumber \\
&-&\Delta_\bot F
\left(\int_{-1}^{\nu}
d\rho D_\bot
-\frac{1}{2} \int_{-1}^{1}
d\mu D_\bot \right)
-\left(\int_{-1}^{\nu}d\rho
D_\bot \Delta_\bot g
-\frac{1}{2} \int_{-1}^{1}d\mu
D_\bot \Delta_\bot g
\right)\Bigg].   \label{R}
\end{eqnarray}
With the latter equation
we can get the formulas of
$\partial{R} / \partial{z}$
and $\Delta_\bot R$
as follows
\begin{eqnarray}
\frac{\partial{R}}
{\partial{z}}=&&
\int_{-1}^{\mu}d\nu
\frac{e^{-M(\nu)}}{D_{\mu \mu}}
\Bigg[\frac{\partial^2{F}}
{\partial{t}
\partial{z}}\nu
+\frac{\partial^2{}}
{\partial{t}\partial{z}}
\int_{-1}^{\nu}gd\rho
+\frac{v}{2}\left(2
\int_{-1}^{\nu}d\rho
\rho \frac{\partial^2{g}}
{\partial{z^2}}
-\int_{-1}^{1}d\mu \mu
\frac{\partial^2{g}}
{\partial{z^2}}\right)
\nonumber\\
&-&\Delta_\bot
\frac{\partial{F}}
{\partial{z}}
\left(\int_{-1}^{\nu}
d\rho D_\bot
-\frac{1}{2} \int_{-1}^{1}
d\mu D_\bot \right)
-\left(\int_{-1}^{\nu}d\rho
D_\bot \Delta_\bot
\frac{\partial{g}}
{\partial{z}}-
\frac{1}{2} \int_{-1}^{1}d\mu
D_\bot \Delta_\bot
\frac{\partial{g}}{\partial{z}}
\right)\Bigg]   \label{dR/dz}
\end{eqnarray}
and
\begin{eqnarray}
\Delta_\bot R
=&&\int_{-1}^{\mu}d\nu
\frac{e^{-M(\nu)}}{D_{\mu \mu}}
\Bigg[\Delta_\bot
\frac{\partial{F}}
{\partial{t}}\nu
+\Delta_\bot\frac{\partial{}}
{\partial{t}}\int_{-1}^{\nu}
gd\rho+\frac{v}{2}
\left(2\int_{-1}^{\nu}d\rho \rho
\Delta_\bot\frac{\partial{g}}
{\partial{z}}
-\int_{-1}^{1}d\mu \mu
\Delta_\bot\frac{\partial{g}}
{\partial{z}}\right)
\nonumber\\
&-&\Delta_\bot^2 F
\left(\int_{-1}^{\nu}d\rho
D_\bot
-\frac{1}{2} \int_{-1}^{1}d\mu
D_\bot \right)
-\left(\int_{-1}^{\nu}d\rho
D_\bot \Delta_\bot^2
g-\frac{1}{2} \int_{-1}^{1}d\mu
D_\bot
\Delta_\bot^2 g\right)\Bigg].
\label{Delta perpendicular R}
\end{eqnarray}

From the latter equations
we can find that the terms
 $\partial^2{g}/\partial{z}^2$,
$\Delta_\bot \partial{g}/\partial{z}$,
$\Delta_\bot \partial{g}/\partial{t}$ and
$\Delta_\bot^2 g$
need to be obtained. Then
inserting the latter equations into
Equation (\ref{Lambda}),
we can obtain the formulas of
$\eta_{1,0,0}$,
$\eta_{0,1,0}$,
$\eta_{0,2,0}$ and
$\eta_{0,0,1}$.

\subsection{The analytical
perpendicular
diffusion coefficient
with $\Lambda(\mu)\ne 0$}
\label{The analytical perpendicular
diffusion coefficient
for Lambdane 0}

From Equation
(\ref{Accurate diffusion
equation-CST})
we can find that
the perpendicular
 diffusion coefficient is
 the corresponding
coefficient of the term
$\Delta_\bot F$,
and the correction
to the perpendicular
diffusion coefficient from
$\Lambda(x,y,z,t)$
is the coefficient
of the term $\Delta_\bot F$ in
$\Lambda(x,y,z,t)$.
As shown in Equation
(\ref{Lambda}),
the term $\Lambda(x,y,z,t)$
is the function of
$\partial{R}/\partial{z}$
and $\Delta_\bot R$.
In the formula of
$\partial{R}/\partial{z}$
(see Equation (\ref{dR/dz})),
the term
$\Delta_\bot F$ does not exist.
In addition, from the formula
of $\Delta_\bot R$
(see Equation
(\ref{Delta perpendicular R}))
we can find that
the term including
 $\Delta_\bot F$ does not exist,
 too.
Therefore, $\Lambda(x,y,z,t)$
does not have the term
$\Delta_\bot F$,
and the correction to
the perpendicular diffusion
coefficient
from $\Lambda(x,y,z,t)$
is zero, i.e.,
$\eta_{0,0,1}=0$.
From Equation (\ref{k001})
we can find that
the perpendicular diffusion
coefficient can be written as
\begin{equation}
\kappa_\bot=\kappa_\bot^a
=\frac{\int_{-1}^{1}d\mu D_\bot
e^{M(\mu)}}{\int_{-1}^{1}
d\mu e^{M(\mu)}}.
\label{k perpendicular
with Lambda}
\end{equation}

In general, the Fokker-Planck
perpendicular diffusion coefficient
$D_\bot$ depends on pith-angle
cosine $\mu$.
By using unified nonlinear transport
theory \citep{Shalchi2010},
\citet{QinandShalchi2014}
developed
a model of $D_\bot (\mu)$
to find
$D_\bot (\mu)\propto |\mu|$.
But this model describes the case
for uniform background magnetic field.
Maybe the weakly nonlinear theory
\citep{ShalchiEA2004}
can be used to derive the formula of
$D_\bot (\mu)$ with
adiabatic focusing effect,
but this procedure is too complicated.
Therefore, so far, no mathematically
tractable theory describing
the relationship of $D_\bot$
to pitch angle cosine $\mu$ and
adiabatic focusing
is obtained
to explore this problem.
In this paper,
we only obtain the latter formula
of the perpendicular diffusion
coefficient with the
adiabatic focusing,
but do not explore it in detail.

\subsection{The analytical
parallel diffusion
coefficient with
$\Lambda(x,y,z,t)\ne 0$}
\label{The analytical
parallel diffusion
coefficient for Lambda ne 0}

As shown in Equation
(\ref{Accurate diffusion
equation-CST}),
coefficient $\eta_{0,2,0}$ is
the correction from
$\Lambda(x,y,z,t)$
to the parallel
diffusion coefficient.
It is obvious that there
is no term
$\partial^2{F}/\partial{z}^2$
in $\Delta_\bot R$
(see Equation
(\ref{Delta perpendicular R})),
but there might be the term
$\partial^2{F}/\partial{z}^2$
in $\partial{R}/\partial{z}$
(see Equation (\ref{dR/dz})).
Therefore, the correction from
$\Lambda(x,y,z,t)$
(Equation (\ref{Lambda}))
to the parallel diffusion
coefficient
should only come from
\begin{equation}
-\frac{v}{2}
\int_{-1}^{1}d\mu \mu e^{M(\mu)}
\left[\frac{\partial{R}}
{\partial{z}}
-\frac{\int_{-1}^{1}d\mu
\frac{\partial{R}}{\partial{z}}
e^{M(\mu)}}
{\int_{-1}^{1}d\mu e^{M(\mu)}}
\right].
\label{Lambda-421}
\end{equation}
From Equation (\ref{dR/dz})
we can find that
only the terms
$\partial^2{g}/\partial{z}^2$,
i.e.,
\begin{equation}
\frac{v}{2}
\int_{-1}^{\mu}d\nu
\frac{e^{-M(\nu)}}
{D_{\mu \mu}}
\left(2\int_{-1}^{\nu}d\rho
\rho \frac{\partial^2{g}}
{\partial{z^2}}
-\int_{-1}^{1}d\mu \mu
\frac{\partial^2{g}}
{\partial{z^2}}\right),
\label{dR/dz-2}
\end{equation}
might have the contribution to
term
$\partial^2{F}/\partial{z}^2$
in $\partial{R}/\partial{z}$.
By operating
$\partial^2{}/\partial{z}^2$ on
the formula of $g(\mu)$
(see Equation (\ref{g})),
we can obtain
\begin{equation}
\frac{\partial^2{g}}
{\partial{z^2}}
=\left(L\frac{\partial^3{F}}
{\partial{z}^3}
-\frac{\partial^2{F}}
{\partial{z}^2}\right)
\left[1-
\frac{2e^{M(\mu)}}{\int_{-1}^{1}
d\mu e^{M(\mu) }}\right]
+e^{M(\mu)}
\left[\frac{\partial^2{R}}
{\partial{z}^2}
-\frac{\int_{-1}^{1}d\mu
e^{M(\mu)}\frac{\partial^2{R}}
{\partial{z}^2}}
{\int_{-1}^{1}d\mu
e^{M(\mu) }}\right].
\label{d^2g/dz^2}
\end{equation}
From Equation (\ref{R}),
$\partial^2{R}/\partial{z}^2$
can be written as
\begin{eqnarray}
\frac{\partial^2{R}}
{\partial{z}^2}=&&
\int_{-1}^{\mu}d\nu
\frac{e^{-M(\nu)}}{D_{\mu \mu}}
\Bigg[\frac{\partial^3{F}}
{\partial{t}
\partial{z}^2}\nu
+\frac{\partial^3{}}
{\partial{t}\partial{z}^2}
\int_{-1}^{\nu}gd\rho
+\frac{v}{2}
\left(2\int_{-1}^{\nu}d\rho
\rho \frac{\partial^3{g}}
{\partial{z^3}}
-\int_{-1}^{1}d\mu \mu
\frac{\partial^3{g}}
{\partial{z^3}}\right)
\nonumber\\
&-&\Delta_\bot
\frac{\partial^2{F}}
{\partial{z}^2}
\left(\int_{-1}^{\nu}
d\rho D_\bot
-\frac{1}{2}
\int_{-1}^{1}d\mu D_\bot \right)
-\left(\int_{-1}^{\nu}
d\rho D_\bot \Delta_\bot
\frac{\partial^2{g}}
{\partial{z}^2}-
\frac{1}{2} \int_{-1}^{1}
d\mu D_\bot
\Delta_\bot
\frac{\partial^2{g}}
{\partial{z}^2}
\right)\Bigg].
\label{d^2R/dz^2}
\end{eqnarray}
Obviously, the term
$\partial^2{F}/\partial{z}^2$
does not exist on the right
hand side of the
latter equation.
Therefore,
we can find that the term
$\partial^2{F}/\partial{z}^2$
on the right hand side of
Equation (\ref{d^2g/dz^2}) is
\begin{equation}
-\frac{\partial^2{F}}
{\partial{z}^2}\left[1-
\frac{2e^{M(\mu)}}
{\int_{-1}^{1}d\mu e^{M(\mu) }}
\right].
\label{d^2g/dz^2-3}
\end{equation}

By replacing
$\partial^2{g}/\partial{z}^2$ in
expression (\ref{dR/dz-2})
with expression
(\ref{d^2g/dz^2-3})
we find the term
$\partial^2{F}/\partial{z}^2$
in $\partial R/\partial z$ as
\begin{equation}
-v\frac{\partial^2{F}}
{\partial{z}^2}
\int_{-1}^{\mu}
d\nu \frac{e^{-M(\nu)}}
{D_{\mu \mu}}
\Bigg\{\frac{\int_{-1}^{1}
d\mu \mu
e^{M(\mu)}}
{\int_{-1}^{1}d\mu e^{M(\mu) }}
+\int_{-1}^{\nu}d\rho
\rho \left[1-
\frac{2e^{M(\mu)}}
{\int_{-1}^{1}d\mu
e^{M(\mu) }}\right]\Bigg\}.
\label{dR/dz-3}
\end{equation}
By substituting
$\partial^2{g}/\partial{z}^2$ in
expression (\ref{Lambda-421})
with the latter expression,
we can obtain the term
$\partial^2{F}/\partial{z}^2$
in $\Lambda(x,y,z,t)$ as
\begin{eqnarray}
-\frac{v^2}{2}
\Bigg\{&&\frac{\int_{-1}^{1} d\mu
\mu e^{M(\mu)}}{\int_{-1}^{1}
d\mu e^{M(\mu)}}\int_{-1}^{1}
d\mu e^{M(\mu)}\int_{-1}^{\mu}
d\nu \frac{e^{-M(\nu)}}
{D_{\mu \mu}}
\Bigg[\int_{-1}^{\nu}
d\rho\rho
\left(1-\frac{2e^{M(\rho)}}
{\int_{-1}^{1}d\mu
e^{M(\mu)}}\right)
+\int_{-1}^{1} d\mu \mu
\frac{e^{M(\mu)}}
{\int_{-1}^{1} d\mu
e^{M(\mu)}}\Bigg] \nonumber\\
&-&\int_{-1}^{1}d\mu
\mu e^{M(\mu)}
\int_{-1}^{\mu}d\nu
\frac{e^{-M(\nu)}}
{D_{\mu \mu}}
\Bigg[\int_{-1}^{\nu}
d\rho\rho
\left(1-\frac{2e^{M(\rho)}}
{\int_{-1}^{1}d\mu
e^{M(\mu)}}\right)
+\int_{-1}^{1} d\mu \mu
\frac{e^{M(\mu)}}
{\int_{-1}^{1} d\mu
e^{M(\mu)}}\Bigg]
\Bigg\}\frac{\partial^2{F}}
{\partial{z}^2},
\label{Lambda-42}
\end{eqnarray}
so the correction coefficient
$\eta_{0,2,0}$
from $\Lambda(\mu)$
to the parallel diffusion
coefficient
can be written as
\begin{eqnarray}
&&\eta_{0,2,0} =\frac{v^2}{2}
\Bigg\{
\int_{-1}^{1}d\mu
\mu e^{M(\mu)}
\int_{-1}^{\mu}d\nu
\frac{e^{-M(\nu)}}{D_{\mu \mu}}
\left[\int_{-1}^{\nu}d\rho\rho
\left(1-\frac{2e^{M(\rho)}}
{\int_{-1}^{1}d\mu
e^{M(\mu)}}\right)
+\int_{-1}^{1} d\mu \mu
\frac{e^{M(\mu)}}
{\int_{-1}^{1} d\mu
e^{M(\mu)}}\right]\nonumber\\
&&-\frac{\int_{-1}^{1}
d\mu \mu
e^{M(\mu)}}{\int_{-1}^{1}
d\mu e^{M(\mu)}}
\int_{-1}^{1}
d\mu e^{M(\mu)}\int_{-1}^{\mu}
d\nu \frac{e^{-M(\nu)}}
{D_{\mu \mu}}
\left[\int_{-1}^{\nu}
d\rho\rho
\left(1-\frac{2e^{M(\rho)}}
{\int_{-1}^{1}d\mu
e^{M(\mu)}}\right)
+\int_{-1}^{1} d\mu \mu
\frac{e^{M(\mu)}}
{\int_{-1}^{1} d\mu
e^{M(\mu)}}\right]
\Bigg\}.  \label{A1}
\end{eqnarray}

Since $\kappa_\parallel^a
=\kappa_{\parallel 0}
+T$ (see Equation (\ref{T}))
and considering
Equation (\ref{k020}),
we can obtain
\begin{equation}
\kappa_\parallel
=\kappa_\parallel^a
+\eta_{0,2,0}=
\kappa_{\parallel 0}
+T+\eta_{0,2,0}.
\label{kappa with A}
\end{equation}
Thus, by considering
$\Lambda(\mu)$
we find a new correction $T'$
to $\kappa_{\parallel 0}$ as
\begin{equation}
\kappa_\parallel
=\kappa_{\parallel 0}+T'
\label{Accurate spatial parallel
diffusion coefficient-1}
\end{equation}
with
\begin{equation}
T'=T+\eta_{0,2,0}.
\label{T'}
\end{equation}
Here, $T$ is the correction
obtained by previous authors
\citep[see,][]{BeeckEA1986,
Litvinenko2012a,ShalchiEA2013,
HeEA2014}, and the formula of
$\eta_{0,2,0}$ is
Equation (\ref{A1}).
From Equations
(\ref{Accurate diffusion
equation-2}) and
(\ref{Accurate diffusion
equation-CST}) we can find
that $\eta_{0,2,0}$
is created by the term
$\Lambda (\vec{x}, t)$
on the right hand side of Equation
(\ref{Accurate diffusion
equation-2}).
The correction coming
from higher-order derivative term might
not be necessarily high-order
small quantity relative
to the correction from other terms.
Therefore, we cannot arbitrarily
neglect any term
in the governing equation
(\ref{Accurate diffusion equation-2}).
In Subsection 4.5, $\eta_{0,2,0}$
will be evaluated and compared with
the correction from other
terms in Equation
(\ref{Accurate diffusion equation-2}).

\subsection{The analytical parallel
streaming coefficient
with $\Lambda(x,y,z,t)\ne 0$}
\label{The analytical parallel
streaming coefficient
for Lambda ne 0}

From Equation
(\ref{Delta perpendicular R})
we can find that the term
$\partial{F}/\partial{z}$
does not
exist in $\Delta_\bot R$,
however, it could exist in
$\partial{R}/\partial{z}$.
Therefore, the correction from
$\Lambda(x,y,z,t)$
to the parallel streaming term
$\partial{F}/\partial{z}$
could only come from the term
\begin{equation}
-\frac{v}{2}\int_{-1}^{1}
d\mu \mu
e^{M(\mu)}
\left[\frac{\partial{R}}
{\partial{z}}
-\frac{\int_{-1}^{1}d\mu
\frac{\partial{R}}{\partial{z}}
e^{M(\mu)}}
{\int_{-1}^{1}d\mu e^{M(\mu)}}
\right].
\label{Lambda-431}
\end{equation}
According to the formula of
$\partial{R}/\partial{z}$
(see Equation (\ref{dR/dz})),
the term
$\partial{F}/\partial{z}$
does not exist in the expression
of $\partial{R}/\partial{z}$.
So that there is no term
$\partial{F}/\partial{z}$
in expression
(\ref{Lambda-431}).
Therefore, there is no correction
from $\Lambda(x,y,z,t)$
to the parallel streaming,  i.e.,
the coefficient $\eta_{0,1,0}$
is equal to 0. In consequence,
from Equation (\ref{k010})
the coefficient of the parallel
streaming is
\begin{equation}
\kappa_1=\kappa_1^a
=v\frac{\int_{-1}^{1}d\mu
\mu e^{M(\mu)}}{\int_{-1}^{1}
d\mu e^{M(\mu)}}.
\end{equation}

\subsection{The analytical coefficient
of the first order time derivative term
with $\Lambda(x,y,z,t)\ne 0$}
\label{The analytical coefficient of
the first order time derivative term
for Lambda ne 0}

From Equation (\ref{R})
we find that
$\Lambda(x,y,z,t)$
is the function of
$\partial{R}/\partial{z}$
and
$\Delta_\bot R$.
Since the term
$\partial{F}/\partial{t}$
does not exist in terms
$\partial{R}/\partial{z}$ and
$\Delta_\bot R$,
from Equation
(\ref{Accurate diffusion
equation-CST}) we can find that
the correction from
$\Lambda(x,y,z,t)$
to the first order
time derivative term
is 0, i.e.,
$\eta_{1,0,0}=0$.
Therefore, from Equation
(\ref{k100})
we can obtain
\begin{equation}
\kappa_t=\kappa_t^a=1.
\end{equation}

\subsection{Evaluating
the correction $T'$
for model
$D_{\mu\mu}=D(1-\mu^2)$}
\label{Calculating the
correction $T'$
for model D}

In the Subsection
\ref{The analytical
parallel diffusion
coefficient for Lambda ne 0},
the parallel
diffusion coefficient
for the case
$\Lambda(x,y,z,t)\ne 0$
has been obtained as
$\kappa_\parallel
=\kappa_{\parallel 0}+T'$
with the correction
$T'=T+\eta_{0,2,0}$.
In Appendix A we find that
the limits of $T$ and
$\eta_{0,2,0}$
all tend to 0 for the limit
$L\rightarrow \infty$.
That is, the correction
$T'$ tends to 0
for the limit
$L\rightarrow \infty$.
Therefore,
$\kappa_\parallel\rightarrow
\kappa_{\kappa 0}$
when the spatially varying
background magnetic field
tends to the uniform one.
In the following we evaluate
the correction $T'$
for the isotropic
model $D_{\mu\mu}=D(1-\mu^2)$.

As shown in \citet{HeEA2014},
for the isotropic pitch-angle
scattering model
$D_{\mu\mu}=D(1-\mu^2)$,
Equation (\ref{M(mu)})
can be simplified as
\begin{equation}
M(\mu)=\xi (\mu+1)
\label{M(mu) for isotropic model}
\end{equation}
with
\begin{equation}
\xi= \frac{v}{2DL}.
\label{xi}
\end{equation}

In Appendix B, by using the
latter simple model
the correction coefficient
$\eta_{0,2,0}$
from $\Lambda(\mu)$
is evaluated.
We can find that
the magnitude of
the lowest order
of the correction
$\eta_{0,2,0}$
is larger than that of
the correction $T$.
If $T$ needs to be considered,
the correction coefficient
$\eta_{0,2,0}$
from $\Lambda(\mu)$
cannot be ignored.

Inserting the quantities
$T$ (Equation
(\ref{expanded T})),
$S$ (Equation (\ref{S}))
and $\eta_{0,2,0}$
(Equation (\ref{A1 with kappa0}))
into the formula of
$\kappa_\parallel$ (Equation
(\ref{Accurate spatial
parallel diffusion
coefficient-1})),
we can obtain
\begin{equation}
\kappa_\parallel
\approx \kappa_{\parallel 0}
\left(1+\frac{2}{15}\xi^2
\right).
\end{equation}
Because $\kappa_\parallel
=\kappa_\parallel^0+T'$,
we can find
\begin{equation}
T'\approx \frac{2}{15}\xi^2
\kappa_{\parallel 0}.
\end{equation}
The lowest order correction
of $T$ obtained by
the previous authors
is equal to
$-\xi^2\kappa_{\parallel 0}/15$
\citep[see,][]{BeeckEA1986,
Litvinenko2012b,
ShalchiEA2013, HeEA2014}.
However, in this paper
the lowest order correction
of $T'$ is equal to
$2\xi^2\kappa_\parallel^0/15$.
Therefore, at least
for the isotropic model
$D_{\mu\mu}=D(1-\mu^2)$
with $\xi\ll 1$,
to consider the adiabatic
focusing effect,
the new correction $T'$
should be used.

\section{SUMMARY AND CONCLUSION}
\label{SUMMARY AND CONCLUSION}

In this paper, by using
the improved perturbation method
of \citet{HeEA2014}
and the iteration process, we explore
the influence of along-field
adiabatic focusing on
energetic charged particle transport.
Starting from the modified
linear Fokker-Planck equation
with the pitch-angle scattering and
perpendicular transport
and adiabatic focusing effect,
we obtain the governing equation of
the isotropic distribution
function $F(\vec{x},t)$
with infinite terms, from which we get
the coefficients of
the spatial parallel
and perpendicular
diffusion, and the
coefficient of parallel
streaming term.
The parallel diffusion
coefficient can be written as
$\kappa_\parallel
=\kappa_{\parallel 0}+T'$,
where $\kappa_{\parallel 0}$
is the parallel diffusion
 coefficient for the uniform
 background magnetic field,
and $T'$ is the correction
to the parallel diffusion coefficient.
We also get
$T'=T+\eta_{0,2,0}$
with  $T$ being the correction
derived in the previous papers
by ignoring the higher-order
derivative terms
in the isotropic
distribution function equation
\citep{BeeckEA1986,
Litvinenko2012a, HeEA2014},
and $\eta_{0,2,0}$ coming
from the higher-order derivative
terms obtained in this paper
but ignored by the previous authors.

Moreover, for the isotropic
pitch-angle scattering model
$D_{\mu\mu}=D(1-\mu^2)$
($D$ is a constant)
with $\xi\ll 1$,
We find that the magnitude of
correction coefficient $\eta_{0,2,0}$
is larger than that
of the correction $T$
obtained in the previous paper.
And the correction
$T'=T+\eta_{0,2,0}$ even has
different sign as
$T$.
In the previous papers,
the higher-order derivative terms
shown by $\Lambda (\vec{x}, t)$ in
the isotropic distribution
function equation
(Equation (\ref{Accurate diffusion
equation-2})) was neglected.
However, in this paper we find
that the higher-order derivative terms
$\Lambda (\vec{x}, t)$
also can make correction
to the parallel
diffusion coefficients, i.e.,
the correction formula $\eta_{0,2,0}$.
Therefore, the correction $T$
obtained in the previous papers is
approximate.
In addition, we find that
the magnitude of
$\eta_{0,2,0}$, which is
to the parallel diffusion coefficient,
is larger than that of $T$
derived by the previous authors.
Therefore, the higher-order
derivative term
in the governing equation
of isotropic distribution
function cannot
be arbitrarily ignored.

In addition, we obtain
the formula of the perpendicular
diffusion coefficient
$\kappa_\bot$.
Since there is no appropriate
theory describing
the relationship of the
Fokker-Planck perpendicular
diffusion coefficient $D_\bot$
to pitch-angle cosine $\mu$
and adiabatic focusing effect,
we do not explore it in detail.
Furthermore,
we find that,
from the terms ignored by
the previous authors
in the governing equation of $F$,
the corrections
to the coefficients of the
spatial perpendicular
diffusion,
the parallel streaming,
and the first order
time derivative term
are equal to 0.

It is noted that $D_{\mu\mu}$
used in computing
$\kappa_\parallel$ in this paper
as well as in previous ones
does not include the
adiabatic focusing effect.
However,
$D_{\mu\mu}$
is corrected
by the adiabatic focusing effect,
which is represented by
the adiabatic focusing
characteristic length $L$,
i.e.,
$D_{\mu\mu}=D_{\mu\mu}(\mu , L)$
\citep{TautzEA2014}.
In fact, pitch-angle
diffusion coefficient $D_{\mu\mu}$ and
perpendicular diffusion coefficient
$D_\bot$ are related to each other
\citep[e.g.,][]{Shalchi2009a}.
Therefore,
$D_\bot$ should also be corrected by
the adiabatic focusing effect,
i.e.,
$D_\bot=D_\bot (\mu,L)$.
So far, no mathematically
tractable theory describing
the relationship of
$D_{\mu\mu}$ and $D_\bot$
to pitch angle cosine $\mu$
and adiabatic focusing
can be used to explore this problem.
In addition, the correction
formula $\eta_{0,2,0}$
(Equation (\ref{A1}))
is too complicated and
it is very difficult to
compute the correction.
Our purpose
is to show
that in order to explore
the correction effect
the higher-order derivative terms
in the governing equation
cannot be neglected.

Compared with the method
used in this paper,
Legendre polynomial
expansions to solve the z-integrated
Fokker-Planck equation
is a more systematic
approach.
But the recursive relation of
different order coefficients
of the expansion series
cannot be obtained.
In addition,
we have not found a method to
transform the z-integrated
Fokker-Planck equation
into another form, from which
one to one relationship
of coefficients of the
expansion series
can be obtained.
In the future we will continue to
investigate this problem.

By using different truncating methods
and transformations,
from the modified
Fokker-Planck equation
one can obtain the
diffusion equation,
the telegraph equation,
or other equations.
Telegraph equation is
also very important.
As another project,
we are exploring
the telegraph equation derived
from the modified
Fokker-Planck equation by employing
the method of \citet{HeEA2014}.

In the future, to
obtain more accurate
analytical formulas of the
spatial parallel and perpendicular
diffusion coefficients, we plan to
get the mathematical
tractable formulas of
$D_{\mu\mu}(L, \mu)$
and $D_\bot(L, \mu)$.
In addition, we plan to
numerically compute
the correction $T'$
and compare it with $T$
for different turbulence
models and
conditions.
In addition, The analysis
to the different
length scales and time scales
in the problem
is also important. We will
explore this problem
by using scale analysis
and dimensional analysis
in the future.

\acknowledgments

We are partly supported by
grants NNSFC 41574172 and NNSFC 41874206.

\appendix
\section{Parallel
diffusion coefficient
$\kappa_\parallel$ for the limit
$L\rightarrow \infty$}
From Equations
(\ref{Accurate spatial parallel
diffusion coefficient-1})
and (\ref{T'}),
the parallel diffusion coefficient
$\kappa_\parallel$
with adiabatic focusing effect
can be written as
$\kappa_\parallel
=\kappa_{\parallel 0}+T'
=\kappa_{\parallel 0}+T+\eta_{0,2,0}$
with the parallel diffusion coefficient
for uniform background field
$\kappa_{\parallel 0}$, the correction
$T$ obtained in previous papers,
and the correction formula $\eta_{0,2,0}$
derived in this paper.
If adiabatic focusing characteristic
length $L$ tends to infinity,
i.e., $\xi$ tends to zero,
the parallel diffusion coefficient
$\kappa_\parallel$
should tend to the uniform field
parallel diffusion coefficient
$\kappa_{\parallel 0}$,
i.e.,
$\lim_{L\to \infty} T'=0$.
Therefore,
$\lim_{L\to \infty} \eta_{0,2,0}=0$
and
$\lim_{L\to \infty} T=0$ .
In the following, we prove
the above limits.

The correction coefficient
$\eta_{0,2,0}$
(see Equation (\ref{A1}))
is the function
of quantity $M(\mu)$
(see Equation (\ref{M(mu)})).
So that we have to explore
$M(\mu)$
for the limit
$L\rightarrow \infty$,
i.e.,
$\lim_{L\to\infty}M(\mu)$.
If the integral
$\int_{-1}^{\mu}d\nu(1-\nu^2)
/D_{\mu\mu}(\nu)$ in $M(\mu)$
(see Equation (\ref{M(mu)}))
is convergent, we can find
$\lim_{L\to\infty}M(\mu)=0$.
Accordingly, the limit
\begin{eqnarray}
\lim_{L\to \infty}
e^{M(\mu)}&=&1,\\
\lim_{L\to \infty}
e^{-M(\mu)}&=&1.
\end{eqnarray}
can be obtained.
Therefore,
we can find
the following relation
\begin{equation}
\lim_{L\to\infty}
\frac{\int_{-1}^1
d\mu\mu e^{M(\mu)}}
{\int_{-1}^1d\mu e^{M(\mu)}}=0
\label{core-1}
\end{equation}
Similarly, the following formula
can be found
\begin{equation}
\lim_{L\to\infty}\int_{-1}^{\mu}
d\nu\nu \left(1
-\frac{2 e^{M(\nu)}}{\int_{-1}^1
d\mu e^{M(\mu)}} \right)
=0.
\label{core-2}
\end{equation}
By inserting Equations
(\ref{core-1})
and (\ref{core-2})
into Equation
(\ref{A1}),
we can easily find that
$\lim_{L\to\infty}\eta_{0,2,0}=0$.

Secondly, we investigate
$\lim_{L\to \infty}
\kappa_\parallel^a$.
We already find that
$\lim_{L\to \infty}M(\mu)
= 0$
in the above paragraph.
Using
$\lim_{L\to\infty}e^{M(\mu)}
= 1+\lim_{L\to\infty}M(\mu)$,
from Equation
(\ref{Approximate spatial parallel
diffusion coefficient})
the following equation can
be obtained
\begin{equation}
\lim_{L\to\infty}
\kappa_\parallel^a
=
vL\frac{\lim_{L\to\infty}
\int_{-1}^{1}d\mu\mu M(\mu)}
{2 +\lim_{L\to\infty}
\int_{-1}^1
d\mu M(\mu)}.
\label{approximate A-1}
\end{equation}
By using integration in parts
and the definition of $M(\mu)$
(see Equation (\ref{M(mu)})),
we can obtain
\begin{equation}
\int_{-1}^{1}d\mu\mu M(\mu)
=\frac{v}{4L}\int_{-1}^1d\mu
\frac{(1-\mu^2)^2}
{D_{\mu\mu}(\mu)}.
\label{int_dmumuM}
\end{equation}
Inserting Equations (\ref{M(mu)})
and (\ref{int_dmumuM})
into Equation
(\ref{approximate A-1}) yields
\begin{equation}
\lim_{L\to\infty}
\kappa_\parallel^a
= \frac{v^2}{4}
\frac{\lim_{L\to\infty}
\int_{-1}^{1}d\mu
(1-\mu^2)^2/D_{\mu\mu}(\mu)}
{2+\lim_{L\to\infty}U/L}
\label{approximate A-1-2}
\end{equation}
with
\begin{equation}
U=\frac{v}{2}\int_{-1}^{1}
d\mu \int_{-1}^{\mu}d\nu
(1-\nu^2)/D_{\mu\mu}(\nu).
\end{equation}
If $U$ is  finite, the following
can be obtained
\begin{equation}
\lim_{L\to\infty}\frac{U}{L}=0,
\end{equation}
and Equation
(\ref{approximate A-1-2})
becomes
\begin{equation}
\lim_{L\to\infty}
\kappa_\parallel^a
= \frac{v^2}{8}
\int_{-1}^{1}d\mu
\frac{(1-\mu^2)^2}{D_{\mu\mu}}.
\label{approximate A-1-3}
\end{equation}
The latter formula is
identical with the
parallel diffusion
coefficient
$\kappa_{\parallel 0}$
for uniform background
magnetic field
\citep{Jokipii1966,
HasselmannEA1968,
Earl1974, Shalchi2006}.
So, $\lim_{L\to\infty} T=0$.
Therefore,
we find $\lim_{L\to\infty}
\kappa_\parallel
=\kappa_{\parallel 0}$,
i.e., $\lim_{L\to\infty}T'= 0$.

\setcounter{equation}{0}  
\section{The correction coefficient
$\eta_{0,2,0}$
from $\Lambda(\mu)$ for the
isotropic model $D_{\mu\mu}=D(1-\mu^2)$}
Here, by employing the isotropic
model $D_{\mu\mu}=D(1-\mu^2)$
we approximately evaluate
$\eta_{0,2,0}$.
After inserting Equation
(\ref{M(mu) for isotropic model})
into Equation (\ref{A1})
we can get
\begin{eqnarray}
\eta_{0,2,0}=&&\frac{v^2}{2}
\Bigg\{
\int_{-1}^{1}d\mu
\mu e^{\xi \mu}
\int_{-1}^{\mu}d\nu
\frac{e^{-\xi\nu}}{D_{\mu \mu}}
\left[
\frac{\nu^2-1}{2}-
2\frac{\int_{-1}^{\nu}
d\rho\rho e^{\xi\rho}}
{\int_{-1}^{1}d\mu
e^{\xi\mu}}+
\frac{\int_{-1}^{1}
d\mu \mu e^{\xi\mu}}
{\int_{-1}^{1} d\mu
e^{\xi\mu}}\right]
\nonumber\\
&-&\frac{\int_{-1}^{1}
d\mu \mu e^{\xi \mu}}
{\int_{-1}^{1}
d\mu e^{\xi \mu}}\int_{-1}^{1}
d\mu e^{\xi \mu}
\int_{-1}^{\mu}d\nu
\frac{e^{-\xi \nu}}
{D_{\mu \mu}}
\left[
\frac{\nu^2-1}{2}-
2\frac{\int_{-1}^{\nu}
d\rho\rho e^{\xi\rho}}
{\int_{-1}^{1}d\mu
e^{\xi\mu}}+
\frac{\int_{-1}^{1}
d\mu \mu e^{\xi\mu}}
{\int_{-1}^{1} d\mu
e^{\xi\mu}}\right]
\Bigg\}.   \label{A1-iso}
\end{eqnarray}
The correction $T$ to the parallel
diffusion coefficient
$\kappa_{\parallel 0}$
has been
obtained in the previous papers
\citep[see,][]{BeeckEA1986,
Litvinenko2012b,
ShalchiEA2013, HeEA2014} as
\begin{eqnarray}
T&=&\kappa_{\parallel 0} S
\label{expanded T}
\end{eqnarray}
with
\begin{eqnarray}
S=-\frac{1}{15}\xi^2
+\frac{2}{315}\xi^4+\cdots.
\label{S}
\end{eqnarray}
To proceed, by employing the
latter formula
we can get
\begin{equation}
\frac{\int_{-1}^{1} d\mu \mu
e^{\xi\mu}}{\int_{-1}^{1}
d\mu e^{\xi\mu}}
=\frac{1+S}{3}\xi.
\label{J1}
\end{equation}
By inputting Equation (\ref{J1}) into
Equation (\ref{A1-iso}),
we can obtain
\begin{equation}
\eta_{0,2,0}=\frac{v^2}{2}
\Bigg[Y_2
-\frac{1+S}{3}\xi Y_1\Bigg]
\label{A1-iso-simpified}
\end{equation}
with
\begin{eqnarray}
Y_1&=&\int_{-1}^{1}
d\mu e^{\xi \mu}
\int_{-1}^{\mu}
d\nu Z(\nu),\label{Y1}\\
Y_2&=&\int_{-1}^{1}d\mu \mu
e^{\xi \mu}
\int_{-1}^{\mu}d\nu  Z(\nu),
\label{Y2}\\
Z(\mu)&=&\frac{e^{-\xi \nu}}
{D_{\mu \mu}}\left[
\frac{\nu^2-1}{2}-
2\frac{\int_{-1}^{\nu}
d\rho\rho
e^{\xi\rho}}
{\int_{-1}^{1}d\mu
e^{\xi\mu}}+
\frac{1+S}{3}\xi\right].
\label{Z(mu)}
\end{eqnarray}
By integration in parts
for Equations (\ref{Y1})
and (\ref{Y2}),
we can obtain
\begin{eqnarray}
Y_1&=&\frac{e^{\xi}}{\xi}
\int_{-1}^1d\mu Z(\mu)
-\frac{1}{\xi}\int_{-1}^1d\mu
e^{\xi\mu} Z(\mu),
\label{Y1-2}\\
Y_2&=&\frac{e^{\xi}}{\xi}
\int_{-1}^1d\mu Z(\mu)
-\frac{1}{\xi}\int_{-1}^1d\mu
e^{\xi\mu} Z(\mu)\mu
-\frac{1}{\xi}Y_1.
\label{Y2-2}
\end{eqnarray}

After inserting Equations
(\ref{Y1-2})
and (\ref{Y2-2}) into Equation
(\ref{A1-iso-simpified}),
we can get
\begin{equation}
\eta_{0,2,0}=\frac{v^2}{2}
\Bigg[e^\xi
\left(\frac{1}{\xi}
-\frac{1}{\xi^2}-\frac{1+S}
{3}\right)
\int_{-1}^1 d\mu Z(\mu)
-\frac{1}{\xi}\int_{-1}^1 d\mu
\mu e^{\mu\xi}Z(\mu)
+\left(\frac{1}{\xi^2}
+\frac{1+S}{3}\right)
\int_{-1}^1 d\mu e^{\mu\xi}
Z(\mu)\Bigg].
\label{A1-iso-single integral}
\end{equation}
For $\xi\ll 1$, by using the
following equations
\begin{eqnarray}
e^\xi&=&1+\xi+\frac{1}{2}\xi^2
+\frac{1}{6}
\xi^3+\frac{1}{24}\xi^4+\cdots,
\label{e1}\\
e^{\mu\xi}&=&1+\mu\xi
+\frac{1}{2}(\mu\xi)^2
+\frac{1}{6}(\mu\xi)^3
+\frac{1}{24}(\mu\xi)^4
+\cdots,
\label{e2}
\end{eqnarray}
we find that Equation
(\ref{A1-iso-single integral})
becomes
\begin{equation}
\eta_{0,2,0}\approx\frac{v^2}{2}
\Bigg[\frac{1}{2}
\int_{-1}^1 d\mu Z(\mu)(1-\mu^2)
+\frac{\xi}{3}\int_{-1}^1
d\mu Z(\mu)\mu
(1-\mu^2)
+\frac{\xi^2}{24}\int_{-1}^1
d\mu Z(\mu)
(1-\mu^2)(3\mu^2-1)\Bigg].
\label{A1-iso-single
integral-approximate}
\end{equation}
Similarly, employing
Equations (\ref{e1})
and (\ref{e2}) with $\xi\ll 1$,
formula (\ref{Z(mu)})
can be simplified as
\begin{equation}
Z(\mu)\approx \frac{1}
{D_{\mu\mu}(\mu)}
\Bigg[-\frac{\mu^3}{3}\xi
+\left(\frac{\mu^2-1}{12}
-\frac{\mu^4-1}{8}
+\frac{\mu^4}{3}\right)
\xi^2\Bigg].
\label{Z(mu)-simplified}
\end{equation}
By combining Equations
(\ref{A1-iso-single
integral-approximate}) and
(\ref{Z(mu)-simplified}),
we can find
\begin{equation}
\eta_{0,2,0}\approx
\frac{v^2}{30D}\xi^2.
\end{equation}
In addition, because of formula
$\kappa_{\parallel 0}=v^2/(6D)$
\citep[see, e.g.,][]
{Shalchi2009a},
 the latter equation becomes
\begin{equation}
\eta_{0,2,0}\approx
\frac{1}{5}\xi^2
\kappa_{\parallel 0}.
\label{A1 with kappa0}
\end{equation}

\end{document}